\documentclass[epj,nopacs]{svjour}

\usepackage[english]{babel}
\usepackage{graphics}
\usepackage{amsmath, amssymb}
\usepackage{amsfonts}
\usepackage{mathtools}
\usepackage{units}
\usepackage{mathrsfs}
\usepackage{color}
\usepackage[square, comma, numbers, super, sort&compress]{natbib}

\newcommand*{\mat}[1]{\boldsymbol{#1}}
\newcommand*{\coord}[1]{\mathbf{#1}}
\newcommand*{\br}{\coord{r}}

\newcommand*{\fv}{\coord{f}}

\DeclarePairedDelimiter{\abs}{\lvert}{\rvert}
\DeclarePairedDelimiterX\braket[2]{\langle}{\rangle}{#1\delimsize\vert#2}
\DeclarePairedDelimiterX\brakket[3]{\langle}{\rangle}{#1\delimsize\vert#2\delimsize\vert#3}
\DeclarePairedDelimiterX\ket[1]{\lvert}{\rangle}{#1}
\DeclarePairedDelimiterX\bra[1]{\langle}{\rvert}{#1}

\newcommand*{\binteg}[3]{\int^{\mathrlap{#3}}_{\mathrlap{#2}}\ud{#1}\,}
\newcommand*{\integ}[1]{\!\int\!\ud{#1}\:}
\newcommand*{\iinteg}[2]{\integ{#1}\!\!\integ{#2}}

\newcommand*{\xc}{\text{xc}}
\newcommand*{\Exc}{E_{\xc}}

\newcommand*{\vxc}{v_{\xc}}
\newcommand*{\vHxc}{v_{\text{H\xc}}}

\newcommand*{\vxcHoleR}{v^{\text{resp}}_{\text{xc,hole}}}
\newcommand*{\vcKin}{v_{\text{c,kin}}}
\newcommand*{\vcKinR}{v^{\text{resp}}_{\text{c,kin}}}

\newcommand*{\vhxcsce}{\vHxc^{\text{SCE}}}
\newcommand*{\vRespSCE}{v_{\text{resp}}^{\text{SCE}}}
\newcommand*{\ron}{\rho}

\newcommand*{\du}{\partial}

\newcommand*{\half}{\frac{1}{2}}

\newcommand*{\isDefinedAs}{=}

\newcommand*{\ud}{\mathrm{d}}

\usepackage{hyperref}

\allowdisplaybreaks[4]

\begin{document}

\title{Sum-rules of the response potential in the strongly-interacting limit of DFT}

\author{Sara Giarrusso\inst{1} \and Paola Gori-Giorgi\inst{1} \and Klaas J. H. Giesbertz\inst{1}%
\thanks{k.j.h.giesbertz@vu.nl}
}

\institute{Department of Theoretical Chemistry and Amsterdam Center for Multiscale Modeling, Faculty of Sciences, Vrije Universiteit, De Boelelaan 1083, 1081HV Amsterdam, The Netherlands}

\date{\today}

\abstract{
The response part of the exchange-correlation potential of Kohn--Sham density functional theory plays a very important role, for example for the calculation of accurate band gaps and excitation energies. Here we analyze this part of the potential in the limit of infinite interaction in density functional theory, showing that in the one-dimensional case it satisfies a very simple sum rule.
}

\maketitle

\section{Introduction}
\label{intro}
The vast majority of electronic structure calculations is performed nowadays with Kohn--Sham (KS) density functional theory (DFT), which is, in principle, an exact reformulation of the many-electron Schrödinger equation. In practice, KS DFT needs to rely on approximations for the exchange-correlation (\xc) functional $\Exc[\ron]$ and its functional derivative with respect to the density $\ron(\br)$, the \xc{} potential $\vxc(\br)$. Finding accurate and versatile \xc{} functionals poses the challenge of transforming a one-body quantity (the density) into the effects of the many-body Coulomb interaction beyond mean field. Most approximations in DFT are based on the results for a uniform and slowly-varying density, for which it is possible to connect the density to the many-body interactions. In recent years, a particular asymptotic case for the exact \xc{} functional, namely its semiclassical limit for a given fixed density,\cite{CotFriKlu-CPAM-13,Lew-arxiv-17,CotFriKlu-arxiv-17} also known as the strictly-correlated electron (SCE) limit,\cite{Sei-PRA-99,SeiGorSav-PRA-07,GorVigSei-JCTC-09,ButDepGor-PRA-12} has emerged as another case in which it is possible to see how the density is mathematically transformed into an electronic interaction.\cite{MalGor-PRL-12,MalMirCreReiGor-PRB-13,MenMalGor-PRB-14} The study of this limit has inspired new approximations, based on the spherically averaged electron density.\cite{WagGor-PRA-14,BahZhoErn-JCP-16,VucGor-JPCL-17} In contrast with the more common LDA or GGA models for the \xc{} hole, these approximations are fully non-local density functionals.

The investigation of the exact properties and features of  $\vxc(\br)$ has always played an important role in understanding and building approximations.\cite{BuiBaeSni-PRA-89,UmrGon-PRA-94,GriLeeBae-JCP-94,FilGonUmr-INC-96,BaeGri-PRA-96,GriLeeBae-JCP-96,BaeGri-JPCA-97,TemMarMai-JCTC-09,HelTokRub-JCP-09, CueAyeSta-JCP-15,CueSta-MP-16,KohPolSta-PCCP-16,HodRamGod-PRB-16,YinBroLopVarGorLor-PRB-16,BenPro-PRA-16,RyaOspSta-JCP-17,HodKraSchGro-JPCL-17} In this work we focus on the \xc{} potential, $\vxc(\br)$, in the SCE limit, and particularly on its response part, which has recently revealed several interesting features.\cite{GiaVucGor-arxiv-18} The response part of the \xc{} potential has been defined \cite{BuiBaeSni-PRA-89,BaeGri-PRA-96,GriLeeBae-JCP-96,BaeGri-JPCA-97} using the formalism of conditional amplitudes,\cite{Hun-IJQC-75-1,Hun-IJQC-75-2,LevPerSah-PRA-84, AbeMaiGro-PRL-10, SchGro-PRL-17} and answers the question:\cite{GriLeeBae-JCP-94,LeeGriBae-ZPD-95,BaeGri-JPCA-97} ``How sensitive is the pair-correlation function on average to local changes in the density?''. 

This piece of the \xc{} potential has been shown to be critical for the correct description of virtual KS orbitals' levels, needed for the calculation of molecular excitation energies in TDDFT,\cite{MeeGriBae-JCTC-14, GriMenBae-JCP-16} as well as for the proper description of electron localization in a dissociating heteronuclear molecule\cite{LeeGriBae-ZPD-95, GriBae-PRA-96, GriLeeBae-JCP-96, BaeGri-JPCA-97, TemMarMai-JCTC-09, HelTokRub-JCP-09, HodKraSchGro-JPCL-17} and for the construction of the Levy-Zahariev potential.\cite{VucLevGor-JCP-17} 

 Here we show that, in cases in which the SCE limit can be solved exactly (one-dimensional and spherically symmetric systems), its response potential satisfies a simple sum rule, see Eqs.~\eqref{eq:sr1},~\eqref{eq:srN} and~\eqref{eq:sr2}.

\section{Strictly-correlated electrons formalism in a nutshell}
\label{sec:1}
Consider the $\lambda$-dependent Hohenberg--Kohn functional within the constrained-search definition\cite{Lev-PNAS-79}
\begin{equation}
F_{\lambda}[\ron]= \min_{\mathclap{\Psi\to\ron}}\brakket{\Psi}{\hat{T} + \lambda \hat{V}_{ee}}{\Psi},
\label{eq:Flambda}
\end{equation}
where $\hat{T}$ is the kinetic energy electronic operator, $\hat{V}_{ee}$ is the electron-electron interaction operator, $\Psi\to\ron$ denotes all the fermionic wavefunctions yielding the density $\ron(\br)$, and $\lambda$ is a coupling constant.
The limit $\lambda\to\infty$ of~\eqref{eq:Flambda} is formally equivalent to a semiclassical limit ($\hbar\to 0$) at fixed one-electron density\cite{CotFriKlu-CPAM-13,Lew-arxiv-17,CotFriKlu-arxiv-17} and it is given by
\begin{equation}
F_{\lambda\to\infty}[\ron]= \lambda\,	V_{ee}^{\text{SCE}}[\ron]+\mathcal{O}\bigl(\sqrt{\lambda}\bigr),
\label{eq:Flambdainfinity}
\end{equation}
where the SCE functional $V_{ee}^{\text{SCE}}[\ron]$ can be written in terms of co-motion functions $\fv_i(\br)$,
\begin{equation}\label{eq:Veesce-reads}
V_{ee}^{\text{SCE}}[\ron]= \half\integ{\br}\ron(\br)\sum_{i=1}^{\mathclap{N-1}} w(\br-\fv_i(\br)),
\end{equation}
with $w(\br) \equiv 1/\abs{\br}$ all along the paper.
The functions  $\fv_i(\br)$ realise the perfect correlation between the $N$ electrons, providing the positions of $N-1$ electrons given the position $\br$ of the first one.\cite{Sei-PRA-99,SeiGorSav-PRA-07}
They are highly non-local functionals of the density satisfying the equation
\begin{align}
\ron\bigl(\fv_i(\br)\bigr)\,\ud\fv_i(\br) &= \ron(\br)\,\ud \br && (i= 1,\dots, N),
\label{eq:differential}
\end{align}
which ensures that the probability of finding one electron at position $\br$ in the volume element $\ud \br$ be the same of finding electron $i$ at position $\fv_i(\br)$ in the volume element $\ud \fv_i(\br)$. They also satisfy cyclic group relations

\begin{equation}
	\label{eq:groupprop}
\begin{aligned}
\fv_0(\br)&\equiv \br,\\
\fv_1(\br)&\equiv \fv(\br),\\
\fv_2(\br)&\equiv \fv\bigl(\fv(\br)\bigr),\\
&\vdotswithin{=} \\
\fv_{N-1}(\br)&\equiv\underbrace{\fv\bigl(\fv(\ldots\fv(\br)\ldots)\bigr)}_{N-1\text{ times}},\\
\fv_N(\br) &\equiv\underbrace{\fv\bigl(\fv(\ldots\fv(\br)\ldots)\bigr)}_{N\text{ times}}=\br.
\end{aligned}
\end{equation}
The functional derivative of the SCE functional, $\vhxcsce(\br) \isDefinedAs \frac{\delta V_{ee}^{\text{SCE}}[\ron]}{\delta \ron(\br)}$, satisfies the following force equation
\begin{align} \label{eq:nablasce}
\mat{\nabla}\vhxcsce(\br)
&= \sum_{i=1}^{\mathclap{N-1}}(\mat{\nabla}w)\bigl(\br - \fv_i(\br)\bigr),
\end{align}
which provides a powerful shortcut to compute $\vhxcsce(\br)$.\cite{MalGor-PRL-12,MalMirCreReiGor-PRB-13,MenMalGor-PRB-14} As usual, the functional derivative $\vhxcsce(\br)$ is defined up to an arbitrary constant from~\eqref{eq:nablasce}, which is usually fixed, for finite systems, by the condition $\vhxcsce(\abs{\br}\to 0)$.

\section{Exchange-correlation response potential}
The exchange-correlation functional can be written as
\begin{equation}
	\label{eq:Excg}
	\Exc[\ron]=\half \iinteg{\br}{\br'} \ron(\br) \ron(\br')\frac{\overline{g}_{\xc}(\br,\br')}{\abs{ \br - \br' }} ,
\end{equation}
where the coupling-constant-averaged pair correlation function $\overline{g}_\text{xc}(\br,\br')$
\begin{equation}\label{eq:pcfcca}
\overline{g}_{\xc}(\br,\br') = \binteg{\lambda}{0}{1} g^{\lambda}_{\xc}(\br,\br') ,
\end{equation}
is obtained by averaging the $g^{\lambda}_{\xc}(\br,\br')$ obtained from the minimizing wavefunction $\Psi^{\lambda}$ in~\eqref{eq:Flambda},
\begin{equation}
	g^{\lambda}_{\xc}(\br,\br') = \frac{P_{2}^{\lambda}(\br,\br')}{\ron(\br) \ron(\br')} - 1,
\end{equation}
with
\begin{equation}\label{eq:P2}
P_{2}^{\lambda}(\br,\br')=N(N-1) \int \abs{\Psi^{\lambda}(\br,\br',\dotsc)}^2 \ud \br_3 \dotsb \ud \br_N.
\end{equation}
The functional derivative of~\eqref{eq:Excg} has then two terms
\begin{equation} \label{eq:vxc-dec-cca}
\vxc(\br)=  \frac{\delta E_{\xc}[\ron]}{\delta\ron(\br)}
= \overline{v}_\text{\xc,hole}(\br) + \overline{v}_\text{resp}(\br),
\end{equation}
where
\begin{equation}
	\label{eq:vxchole}
\overline{v}_\text{\xc,hole}(\br) = \integ{\br'} \ron(\br') \frac{\overline{g}_{\xc}(\br,\br')}{\abs{ \br - \br' }} ,
\end{equation}
and
\begin{equation}
\overline{v}_\text{resp}(\br)= \half \iinteg{\br'}{\br''}  \frac{\ron(\br')\ron(\br'')}{\abs{ \br - \br' }} \frac{\delta \overline{g}_{\xc}(\br', \br'')}{\delta\ron(\br)} .
\label{vrespbar}
\end{equation}
A different definition of response potential, which results from taking the functional derivative of the \xc{} energy expressed as a sum of kinetic and Coulomb interaction terms, is also documented in the literature\cite{ GriLeeBae-JCP-94, GriBae-PRA-96, BaeGri-JPCA-97}
\begin{equation}\label{vresp}
v_\text{resp}(\br) = \vcKinR(\br) + \vxcHoleR(\br),
\end{equation}
where $\vcKin(\br) $ is a kinetic correlation energy density, such that
$
\integ{\br} \vcKin(\br) \ron(\br) = T[\ron]-T_s[\ron],
$
and its response part is defined as
\begin{equation}\label{eq:vckinresp}
\vcKinR(\br) = \integ{\br'} \frac{\delta \vcKin(\br')}{\delta \ron(\br)} \ron(\br')
\end{equation}
and where 
\begin{equation}\label{eq:vxcholeresp}
\vxcHoleR(\br) = \half \iinteg{\br'}{\br''}  \frac{\ron(\br')\ron(\br'')}{\abs{ \br - \br' }} \frac{\delta g_{\xc}(\br', \br'')}{\delta\ron(\br)},
\end{equation}
with $g_{\xc} = g_{\xc}^{\lambda =1}$.
For an extensive overview of the origin and implications of the two definitions, the reader is referred to Ref.~\citenum{GiaVucGor-arxiv-18}.

Similarly, we can write the SCE \xc{} energy functional as
\begin{equation}
	\Exc^{\text{SCE}}[\ron]= \half \iinteg{\br}{\br'} \ron(\br) \ron(\br') \frac{g_{\xc}^{\infty}(\br , \br')}{\abs{ \br - \br' }} ,
\end{equation}
where it should be noted that, since the kinetic component in the strongly-interacting limit is subleading, 
\begin{equation}
g_\text{xc}^{\infty}(\br,\br') \sim \frac{1}{\lambda} \binteg{\lambda'}{0}{\lambda} g^{\lambda'}_{\xc}(\br,\br') \quad \lambda \rightarrow \infty
\end{equation}
and the expressions of the SCE \xc{} energy in terms of coupling-constant averaged or kinetic and Coulomb interaction quantities coincide (again, see Ref.~\citenum{GiaVucGor-arxiv-18} for an in-depth discussion). Thus, the distinction between the two possible definitions of the response part disappears, leading to a univocal definition of the response potential in this limit as
\begin{equation} \label{eq:def-vrespSCE}
\vRespSCE(\br)= \half \iinteg{\br}{\br'}  \frac{\ron(\br') \ron(\br'')}{\abs{ \br' - \br'' }} \frac{\delta g_{\xc}^{\infty}(\br' , \br'')}{\delta\ron(\br)} .
\end{equation}
Finally, in Ref.~\citenum{GiaVucGor-arxiv-18} it has been shown that Eq.~\ref{eq:def-vrespSCE} can be cast into the much more handy expression
\begin{equation}\label{eq:vrespsce}
\vRespSCE(\br) = \vhxcsce(\br) - \sum_{i=1}^{\mathclap{N-1}}w\bigl(\br - \fv_i(\br)\bigr).
\end{equation}
From this equation it is clear that $\vRespSCE$ inherits the asymptotic value of $\vhxcsce$.

\section{Sum-rule of the SCE response potential}
We can use~\eqref{eq:nablasce} and~\eqref{eq:vrespsce} to derive the following expression for the gradient of the SCE response potential
\begin{align}\label{eq:nablavrespsce}
\mat{\nabla}\vRespSCE(\br)
={}& \sum_{i=1}^{\mathclap{N-1}}(\mat{\nabla}w)\bigl(\br - \fv_i(\br)\bigr) \notag \\
&{}- \sum_{i=1}^{\mathclap{N-1}}(\mat{\nabla}w)\bigl(\br - \fv_i(\br)\bigr)
\bigl(\mat{1} - \mat{\nabla}\fv_i(\br)\bigr) \notag \\
{}={}& \sum_{i=1}^{\mathclap{N-1}}
(\mat{\nabla}w)\bigl(\br - \fv_i(\br)\bigr) \cdot \mat{\nabla}\fv_i(\br) ,
\end{align}
where the dot product is taken over the components of the co-motion function. To clarify, let us work out the expression per component in Cartesian coordinates
\begin{equation}
\du_{\mu} \vRespSCE(\br)
= -\sum_{\nu=1}^D\sum_{i=1}^{\mathclap{N-1}}
\frac{\br_{\nu} -\fv_{i,\nu}(\br)}{\abs{\br - \fv_i (\br)}^3}\du_{\mu}\fv_{i,\nu}(\br) .
\end{equation}
For the case $D=1$, the response potential can now directly be calculated as an integral.
In the following sections we are going to prove the exact behaviour of the integral of the SCE response potential corresponding to an $N$-electron 1D density and the one corresponding to a spherical two-electron density.
These are the two cases in which the co-motion functions have an analytic expression in terms of the density.\cite{Sei-PRA-99,SeiGorSav-PRA-07,ButDepGor-PRA-12,ColDepDim-CJM-15}

\subsection{Sum-rule of the SCE response potential for a 1D density}
The sum-rule of the SCE response function in 1D (for Coulomb interaction) relates the integral over the response function to the number of electrons. To illustrate the idea, we will first consider the simplest situation: a symmetric 2-electron density. Next we release the symmetry constraint and then generalise to an arbitrary amount of particles. But first, we need an explicit expression for the co-motion functions.\cite{Sei-PRA-99,ColDepDim-CJM-15}

Let us define the cumulant function, $N_e(x)$, for a 1D density
\begin{equation}
N_e(x) \isDefinedAs \binteg{y}{-\infty}{x} \ron(y) .
\end{equation}
We see that the cumulant evaluated at infinity yields the number of electrons, $N_e(x \to \infty) = N $. Since $N_e(x)$ is obviously a monotone function, its inverse $N_e^{-1}(\nu)$ can be defined on the domain $(0, N)$.
We also define the distances, $a_i$, such that the cumulant evaluated in these points give an integer number of electron, $ N_e(a_i)=i$.

By requiring that the co-motion functions fulfill~\eqref{eq:differential} for a 1D density one finds~\cite{Sei-PRA-99,ColDepDim-CJM-15}
\begin{equation}\label{eq:cf1d}
f_i(x) = \begin{cases*}
N_e^{-1}\bigl(N_e(x) + i\bigr)		&for $x < \bar{a}_i$ \\
N_e^{-1}\bigl(N_e(x) + i - N\bigr)	&for $x > \bar{a}_i$ ,
\end{cases*}
\end{equation}
where $\bar{a}_i \isDefinedAs a_{N-i} = N_e^{-1}(N - i)$. From this explicit form of the co-motion functions, it is clear that
\begin{equation}
\binteg{y}{x}{f_i(x)}\ron(y) = i .
\end{equation}
This equation means that the position $f_i$ is exactly $i$ electron(s) to the right. This picture even holds, if one regards the system to be periodic, so that particles disappearing at $+\infty$ reappear at $-\infty$.

\subsubsection{Symmetric two-electron density in 1D}\label{sec:s2-1d}
In the case of a symmetric 1D density with only two electrons, we see that $a_1 = 0$  and the SCE response potential can be expressed as
\begin{equation}\label{eq:vresces2-1d}
\vRespSCE(x)
= \begin{dcases}
\binteg{y}{-\infty}{x}\frac{f'(y)}{\bigl(y - f(y)\bigr)^2}	&(x \leq 0) \\
\binteg{y}{x}{\infty}\frac{f'(y)}{\bigl(y - f(y)\bigr)^2}	&(x \geq 0) ,
\end{dcases}
\end{equation}
where we used that the potential can be obtained by integrating from either side, as the response potential is symmetric.

Let us only consider the negative side of the SCE response potential. By interchanging the order of integration, we find for the integral over the response function
\begin{subequations}
\begin{equation}
\binteg{x}{-\infty}{0}\vRespSCE(x)
= -\binteg{y}{-\infty}{0}\frac{y f'(y)}{\bigl(y - f(y)\bigr)^2} .
\end{equation}
We can also make a change of variables $u = -f(y)$, keeping in mind that, due to the property in~\eqref{eq:groupprop}, $ f^{-1}(x) = f(x)$
\begin{equation}
\binteg{x}{-\infty}{0}\vRespSCE(x)
= \binteg{u}{-\infty}{0}\frac{f(u)}{\bigl(f(u) - u\bigr)^2} .
\end{equation}
\end{subequations}
We can combine these two expressions to write the integral over the SCE response function as
\begin{equation}
\binteg{x}{-\infty}{0}\vRespSCE(x)
= -\half\binteg{u}{-\infty}{0}\frac{y f'(y) - f(y)}{\bigl(y - f(y)\bigr)^2}
= \half .
\end{equation}
As the SCE response potential is a symmetric function, we find that the integral over the real line gives 
\begin{equation}\label{eq:sr1}
\binteg{x}{-\infty}{\infty}\vRespSCE(x) = 1 .
\end{equation}

\subsubsection{General two-electron density in 1D}
In the case of a non-symmetric density we now have almost the same expression for the SCE response potential as in~\eqref{eq:vresces2-1d}, except that we need to cut it at $a_1 = N_e^{-1}(1)$ instead of zero
\begin{equation}
\vRespSCE(x)
= \begin{dcases}
\binteg{y}{-\infty}{x}\frac{f'(y)}{\bigl(y - f(y)\bigr)^2}	&(x \leq a_1) \\
\binteg{y}{x}{\infty}\frac{f'(y)}{\bigl(y - f(y)\bigr)^2}	&(x \geq a_1) ,
\end{dcases}
\end{equation}
where we used again that it does not matter from which side we do the integration. Though physically reasonable, we lack the symmetry argument and have provided an explicit derivation in Appendix~\ref{ap:eqAsymp}.

Now let us first consider the integral over $(-\infty,a_1)$. Again by changing the order of integration, we find
\begin{subequations}
\begin{equation}
\binteg{x}{-\infty}{a_1}\vRespSCE(x)
= a_1v(a_1) - \binteg{y}{-\infty}{a_1}\frac{y f'(y)}{\bigl(y - f(y)\bigr)^2} .
\end{equation}
The integral over $(a_1,\infty)$ yields
\begin{equation}
\binteg{x}{a_1}{\infty}\vRespSCE(x)
= \binteg{y}{a_1}{\infty}\frac{y f'(y)}{\bigl(y - f(y)\bigr)^2} - a_1v(a_1) ,
\end{equation}
\end{subequations}
so the full integral over the response function becomes
\begin{subequations}
\begin{equation}\label{eq:expr1}
\binteg{x}{-\infty}{\infty}\vRespSCE(x)
= \left(\binteg{y}{a_1}{\infty} - \binteg{y}{-\infty}{a_1}\right)\frac{y f'(y)}{\bigl(y - f(y)\bigr)^2} .
\end{equation}
Now making the transformation $u = f(y)$, we obtain the following alternative expression
\begin{equation}\label{eq:expr2}
\binteg{x}{-\infty}{\infty}\vRespSCE(x)
= \left(\binteg{u}{-\infty}{a_1} - \binteg{u}{a_1}{\infty}\right)\frac{f(u)}{\bigl(f(u) - u\bigr)^2} .
\end{equation}
\end{subequations}
If we now take the average over~\eqref{eq:expr1} and~\eqref{eq:expr2}, we find again that the full integral yields
\begin{multline}
\binteg{x}{-\infty}{\infty}\vRespSCE(x) \\
= \half\left(\binteg{y}{a_1}{\infty} - \binteg{y}{-\infty}{a_1}\right)
\frac{y f'(y) - f(y)}{\bigl(y - f(y)\bigr)^2} = 1 .
\end{multline}

\subsubsection{Arbitrary amount of electrons in 1D}
As the number of electrons exceeds two, we deal with a set of co-motion functions. As we do not have $f = f^{-1}$ anymore, we need to find the inverses of each co-motion functions  in~\eqref{eq:cf1d}. These are  
\begin{align}
f^{-1}_i(x) = \begin{cases*}
N_e^{-1}\bigl(N_e(x) - i\bigr)		&for $x < a_i$ \\
N_e^{-1}\bigl(N_e(x) - i + N\bigr)	&for $x > a_i$ ,
\end{cases*}
\end{align}
where we see that, as expected, they are also co-motion functions, $f^{-1}_i = f_{N-i}$.
Now let us consider the general SCE response potential in 1D
\begin{multline}
\vRespSCE(x)
= \sum_{i=1}^{\mathclap{N-1}}\biggl(\theta(\bar{a}_i - x)\binteg{y}{-\infty}{x} \\
{} + \theta(x - \bar{a}_i)\binteg{y}{x}{\infty}\biggr)
\frac{f_i'(y)}{\bigl(y - f_i(y)\bigr)^2} ,
\end{multline}
where the expression for $x > \bar{a}_i$ is again justified by~\eqref{eq:vRespConsist} for each $f_i$.

By interchanging the integration again, the integral over the SCE response potential can be expressed as
\begin{subequations}
\begin{multline}
\binteg{x}{-\infty}{\infty}\vRespSCE(x) \\
= \sum_{i=1}^{\mathclap{N-1}}\left(\binteg{y}{\bar{a}_i}{\infty} - \binteg{y}{-\infty}{\bar{a}_i}\right)
\frac{y f_i'(y)}{\bigl(y - f_i(y)\bigr)^2} . 
\end{multline}
Making the variable transformation $u = f_i(y)$, we find
\begin{multline}
\binteg{x}{-\infty}{\infty}\vRespSCE(x)
= \sum_{i=1}^{\mathclap{N-1}}\left(\binteg{u}{-\infty}{a_i} - \binteg{u}{a_i}{\infty}\right)
\frac{f^{-1}_i(u)}{\bigl(f^{-1}_i(u) - u\bigr)^2} \\
= \sum_{i=1}^{\mathclap{N-1}}
\left(\binteg{u}{-\infty}{\bar{a}_{N-i}} - \binteg{u}{\bar{a}_{N-i}}{\infty}\right)
\frac{f_{N-i}(u)}{\bigl(f_{N-i}(u) - u\bigr)^2}  .
\end{multline}
\end{subequations}
As the summation can be done in any order, we can combine it with the previous expression to find
\begin{align}\label{eq:srN}
\binteg{x}{-\infty}{\infty}\vRespSCE(x)
&= \half\sum_{i=1}^{\mathclap{N-1}}
\left(\binteg{y}{\bar{a}_i}{\infty} - \binteg{y}{-\infty}{\bar{a}_i}\right)
\frac{y f_i'(y) - f_i(y)}{\bigl(y - f_i(y)\bigr)^2} \notag \\*
&= N - 1 ,
\end{align}
which proves the interesting property that the integral over the SCE response potential for an $N$-electron 1D density (and Coulomb interaction) gives $N-1$.

\subsection{Sum-rule of the SCE response potential for spherical two-electron densities}

In the case of a 3D spherical density, the spherical volume element now needs to be included in the cumulant
\begin{equation}\label{eq:nerad}
 N_e(r)= \binteg{x}{0}{r} 4\pi x^2 \ron(x) .
\end{equation} 
Although, an ansatz has been proposed for the radial part of the co-motion functions for any arbitrary amount of electrons~\cite{SeiGorSav-PRA-07}, this ansatz has been proven to be exact only for $N = 2$. For $N > 2$ counterexamples have been found where these co-motion functions are not truly optimal~\cite{SeiDiMGerNenGieGor-arxiv-17} but they do still satisfy~\eqref{eq:nablasce}.
As we need an explicit form of the interaction in terms of the radial coordinates, we will limit ourselves to the two-electron case. The radial co-motion function can be worked out as~\cite{SeiGorSav-PRA-07}
\begin{equation}
f(r)= N_e^{-1}\bigl(2-N_e(r)\bigr) .
\end{equation}
The differential equation for the response potential~\eqref{eq:nablavrespsce} in the spherical two-electron case is readily worked out as
\begin{equation}
\frac{\ud}{\ud r} \vRespSCE (r) = \frac{f'(r)}{(r + f(r))^2} ,
\end{equation}
where $\abs{\br - \fv (\br)} = r + f(r)$, since the electrons need to be situated opposite to each other with respect to the origin to minimise their repulsion.
Using the standard gauge again, we have
\begin{equation}
\vRespSCE(s) = -\binteg{r}{s}{\infty} \frac{\ud}{\ud r} \vRespSCE(r).
\end{equation}
We now evaluate the following integral over the response potential
\begin{subequations}
\begin{equation}\label{eq:target}
\binteg{s}{0}{\infty} \vRespSCE(s)
=-\binteg{r}{0}{\infty}  \frac{r  f'(r)}{(r + f(r))^2}.
\end{equation}
Finally, as seen in the 1D case, via the usual transformation $u=f(r) $, we write equivalently the last expression in the above equations as
\begin{equation}
\binteg{s}{0}{\infty} \vRespSCE(s) = \binteg {u}{0}{\infty}   \frac{f(u)}{(u + f(u))^2}. 
\end{equation}
\end{subequations}
By averaging between the two, one obtains that the integral over the positive real line of the SCE response potential for a spherical two-electron density gives
\begin{equation}\label{eq:sr2}
\binteg {r}{0}{\infty} \vRespSCE (r)= \frac{1}{2} \binteg {u}{0}{\infty}  \frac{f(u) - u f'(u)}{(u + f(u))^2}
= \half .
\end{equation}

\section{Concluding remarks}
We have analyzed the SCE response potential and shown that it satisfies a simple sum rule in the one-dimensional and in the $N=2$ spherically symmetric case. This latter case might be a special one, as it is mathematically equivalent to a 1D case, thus requiring further investigation for a generalisation to 3D systems.
Additional investigations are also required whether the sum-rules could also be established for the physical interacting system, either with or without the kinetic part of the response potential.

The response part of the exchange-correlation potential is exactly the part that is less well approximated by standard functionals, which usually provide a decent approximation for the \xc{} hole part of~\eqref{eq:vxchole}. Its exact properties in the extremely correlated scenario provided by the SCE potential may help on building new approximations to this term.

\section*{Acknowledgements}
It is a pleasure to dedicate this paper to Hardy Gross, who has always enjoyed discovering new exact properties in (TD)DFT.

Financial support was provided by the European Research Council under H2020/ERC Consolidator Grant corr-DFT [Grant Number 648932].

\section*{Author contribution statement}
All the authors were involved in the preparation of the manuscript. 
All the authors have read and approved the final manuscript.

\appendix

\section{Proof of equal asymptotics}
\label{ap:eqAsymp}
Here we show explicitly that $\vRespSCE(-\infty) = \vRespSCE(+\infty)$ for $N=2$. To do this, we will work in the cumulant coordinate. To this purpose let us work out the following identities
\begin{subequations}
\begin{align}
\ud N_e(x) &= \ron(x)\,\ud x , \\
\frac{\ud N_e^{-1}(\nu)}{\ud \nu} &= \frac{1}{\ron\bigl(N_e^{-1}(\nu)\bigr)} .
\end{align}
\end{subequations}
Now we work out the response function at $x = +\infty$ by performing the full integral
\begin{align}\label{eq:vRespConsist}
\vRespSCE(\infty)
={}& \binteg{y}{-\infty}{a_1}\frac{f'(y)}{\bigl(y - f(y)\bigr)^2} -
\binteg{y}{a_1}{\infty}\frac{f'(y)}{\bigl(y - f(y)\bigr)^2} \notag \\
={}& \binteg{\nu}{0}{1}
\frac{1/\ron\bigl(N_e^{-1}(\nu + 1)\bigr)}{\bigl(N_e^{-1}(\nu) - N_e^{-1}(\nu + 1)\bigr)^2} + \notag \\
&{}-\binteg{\nu}{1}{2}
\frac{1/\ron\bigl(N_e^{-1}(\nu - 1)\bigr)}{\bigl(N_e^{-1}(\nu) - N_e^{-1}(\nu - 1)\bigr)^2}  \notag \\
={}& \binteg{\nu}{0}{1}\frac{1}{\bigl(N_e^{-1}(\nu) - N_e^{-1}(\nu + 1)\bigr)^2} \notag \\
&{}\times \left(\frac{1}{\ron\bigl(N_e^{-1}(\nu + 1)\bigr)} - \frac{1}{\ron\bigl(N_e^{-1}(\nu)\bigr)}\right) \notag \\
={}& \left[\frac{1}{N_e^{-1}(\nu) - N_e^{-1}(\nu + 1)}\right]_0^1 = 0 .
\end{align}
This explicit demonstration trivially generalises to general $N$ by including a summation over the contributions from each particle.

\bibliographystyle{unsrtnat}
\bibliography{mybib}

\end{document}